\documentclass[manuscript]{acmart}
\usepackage[percent]{overpic}

\begin{document}

\title{Optimization and Portability of a Fusion OpenACC-based FORTRAN HPC Code from NVIDIA to AMD GPUs}

\author{Igor Sfiligoi}
\orcid{0002-9308-5327}
\affiliation{%
  \institution{University of California San Diego}
  \city{La Jolla}
  \state{CA}
  \country{USA} 
}
\email{isfiligoi@sdsc.edu}

\author{Emily A. Belli}
\affiliation{%
  \institution{General Atomics}
  \city{La Jolla}
  \state{CA}
  \country{USA} 
}
\email{bellie@fusion.gat.com}

\author{Jeff Candy}
\affiliation{%
  \institution{General Atomics}
  \city{La Jolla}
  \state{CA}
  \country{USA} 
}
\email{candy@fusion.gat.com}

\author{Reuben D. Budiardja}
\affiliation{%
  \institution{Oak Ridge National Laboratory}
  \city{Oak Ridge}
  \state{TN}
  \country{USA} 
}
\email{reubendb@ornl.gov}

\newcommand{\shz}{\texttt{sh03b}}
\newcommand{\emz}{\texttt{em04b}}


\begin{abstract}
  NVIDIA has been the main provider of GPU hardware in HPC systems for over a decade. Most applications that benefit from GPUs have thus been developed and optimized for the NVIDIA software stack.
  Recent exascale HPC systems are, however, introducing GPUs from other vendors, e.g. with the AMD GPU-based OLCF Frontier system just becoming available. AMD GPUs cannot be directly accessed using the NVIDIA software stack, and require a porting effort by the application developers.
  This paper provides an overview of our experience porting and optimizing the CGYRO code, a widely-used fusion simulation tool based on FORTRAN with OpenACC-based GPU acceleration. While the porting from the NVIDIA compilers was relatively straightforward using the CRAY compilers on the AMD systems, the performance optimization required more fine-tuning. In the optimization effort, we uncovered code sections that
  had performed well on NVIDIA GPUs, but were unexpectedly slow on AMD GPUs. After AMD-targeted code optimizations, performance on AMD GPUs has increased to meet our expectations. Modest speed improvements were also seen on NVIDIA GPUs, which was an unexpected benefit of this exercise.
\end{abstract}




\maketitle

\section{Introduction}

The CGYRO code \cite{candy:2016} is a widely-used, community fusion simulation tool that solves the five-dimensional gyrokinetic-Maxwell equations describing the evolution of plasma microturbulence in magnetic fusion devices.  The code uses an Eulerian approach and is spectral/pseudo-spectral in four of the five phase space dimensions.  CGYRO is highly optimized for so-called \textit{multiscale} turbulence simulation that couples the space and time scales between slow, large-scale motion of the hydrogenic fuel ions and fast, small-scale motion of the electrons.  These scales are separated by nearly two orders of magnitude.  Thus, multiscale simulation requires high mesh resolution in multiple dimensions.  The spatial discretization and array distribution schemes in CGYRO were designed to target scalability on modern but disparate architectures, including next-generation HPC systems that use multicore and GPU-accelerated hardware \cite{candy:2019,sfiligoi:2022}.  This paper describes the porting and optimization effort of CGYRO from HPC systems based on NVIDIA GPUs to AMD GPU-based systems, using the NERSC Perlmutter and OLCF Frontier machines for performance comparisons.

\section{The porting effort}

CGYRO was developed years before GPU computing became the mainstream of HPC systems, and was initially based on the OpenMP paradigm for in-process parallelization. When the first NVIDIA GPUs became available, OpenMP did not have support for GPU offload, so the authors used OpenACC for GPU-based parallelization. NVIDIA has been one of the main drivers of OpenACC and has had great support for it in its software stack ever since.  Unfortunately, no other GPU vendor presently supports OpenACC in their native software stack. But the HPE CRAY FORTRAN compiler does provide OpenACC GPU offloading support, so we defaulted to the CRAY software stack when compiling CGYRO on the new AMD GPU-based OLCF Frontier system.

Compiling the FORTRAN OpenACC-based code using the CRAY compiler for AMD GPUs was relatively straightforward. There were a few places where the compiler indicated that the existing OpenACC directives were missing required fields, even though we are reasonably confident that the OpenACC specification lists them as optional and the NVIDIA compiler did not have an issue. Nevertheless, the error messages were clear and the required changes minimal. Most notably though, while most of the CGYRO code is custom, it does rely heavily on external FFT libraries. On CPU-only architectures CGYRO uses FFTW, while on NVIDIA GPU-based hybrid architectures cuFFT is used. Neither library can be used with AMD GPUs, but AMD provides a hipFFT library that uses AMD's rocFFT as the backend.  The interfaces of the cuFFT and hipFFT libraries are very similar with respect to the type and function names, so adding support for hipFFT in CGYRO was mostly trivial.  It should be noted that hipFFT supports several backends, including cuFTT, and could thus also be used to access NVIDIA GPUs, but we do not use it on these systems.

\section{The optimization effort}

Upon successful code compilation and regression testing on the AMD GPU-based systems, an extensive optimization effort was undertaken.  The optimization focused on comparing the relative performance on AMD MI250X GPU-based nodes on the OLCF Frontier system with the performance on NVIDIA A100 GPU-based nodes on the NERSC Perlmutter. Two cases were chosen for the benchmark: \shz, a modest-sized but physically representative simulation input, and \emz, a large-sized \textit{multiscale} simulation input.  The dimensions of each problem are given in Table~\ref{table:test_cases} for comparison. On NVIDIA GPUs, a single MPI process per physical GPU was used, while on the AMD GPUs, 2 MPI processes per physical GPU were used since each MI250X presents itself to the system as two logical GPUs. In this section we concentrate on results for pure compute performance, while communication performance is discussed in the next section.  The main CGYRO kernels that will be discussed are outlined in Table \ref{table:kernels}.

\begin{table}[h]
\centering
\begin{tabular}{|c|c|}
\hline
   \shz & $\left( N_x,N_y,N_\theta, N_\xi, N_v \right) = (480,48,32,24,8)$  \\
   \hline
   \emz & $\left( N_x,N_y,N_\theta, N_\xi, N_v \right) = (1344,288,24,18,8)$ \\
     \hline
\end{tabular}
\caption{5D resolutions of CGYRO benchmarking test cases for each species.  Both test cases simulated 3 species.  The parameter dimensions are defined in Ref.~\cite{candy:2016}.}
\label{table:test_cases}
\end{table}

\begin{table}[h]
\centering
\begin{tabular}{|c|c|c|}
\hline
Kernel & Dominant Operation & Communication \\
\hline
   field & loop & MPI\_ALLREDUCE 
  \\ \hline
  stream & loop & MPI\_ALLREDUCE 
  \\ \hline
  shear & loop & MPI\_ALLREDUCE 
  \\ \hline
   nonlinear & 2D FFT & MPI\_ALLTOALL \\ \hline
   collision & dense matrix-vector mult & MPI\_ALLTOALL \\
     \hline
\end{tabular}
\caption{Table of main CGYRO kernels and corresponding dominant compute and communication operations.  The kernels are explicitly defined in Ref.~\cite{candy:2019}.}
\label{table:kernels}
\end{table}

The initial benchmark results for the modest-sized \shz\ test case were relatively disappointing.  For this benchmark, CGYRO was run using 24 GPUs on both systems.  In the initial benchmark, we found that CGYRO was significantly slower overall on the AMD GPU-based system than on the NVIDIA GPU-based system. This was surprising since the theoretical specifications for the AMD MI250X are superior to those for the A100 GPU. However, looking at the individual timings for the various logical steps in the CGYRO code, we did observe that at least one code section -- the \textit{collision} kernel -- was indeed faster on the AMD GPUs.  So we decided to explore code optimization opportunities.  

\begin{figure}[h]
  \centering
  \begin{overpic}[width=\linewidth]{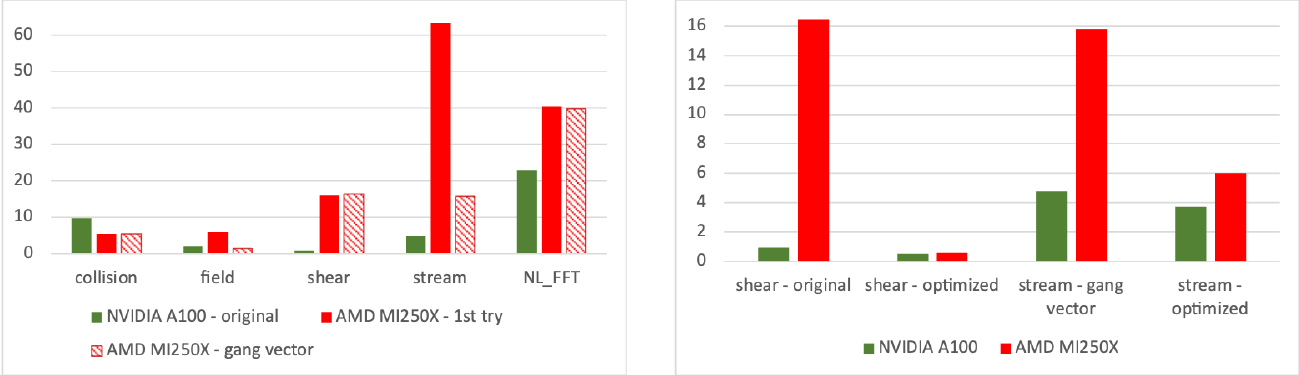}
  \put (1,2){(a)}
  \put (53,2){(b)}
  \end{overpic}
  \caption{CGYRO \shz\ benchmark results for wallclock time (s) comparing performance of various kernel operations on the NVIDIA A100 (green) and AMD MI250X (red) GPU-based systems.  On both systems, CGYRO was run using 24 GPUs. (a) The left plot shows the results upon initial porting (solid red bars) and after optimization steps on the AMD GPU system for explicit prescription of gang vector parallelization (hatched red bars). This most notably sped-up the \textit{field} and \textit{stream} kernels. (b) The right plot shows the results on both systems after targeted optimizations for the \textit{shear} and \textit{stream} kernels.}
  \Description{Histograms with benchmark results.}
  \label{fig1_2}
\end{figure}

We started with comparing the performance of the CGYRO \textit{collision} and \textit{field} kernels.  While both kernels contain mostly simple loops, we observed that they were performing differently on the AMD GPUs. Upon further investigation, we discovered that while we were explicitly directing the compiler to use the OpenACC \textbf{gang vector} parallelization in \textit{collision}, we did not prescribe it in \textit{field}. Adding the directives explicitly significantly sped up the \textit{field} kernel.  This indicated that the CRAY compiler was less capable of automatically choosing the ideal parallelization strategy compared with the NVIDIA compiler. We subsequently added the directives in all suitable locations. Ultimately, this resulted in significant speedups in both the \textit{field} and \textit{stream} kernels, as shown in Figure \ref{fig1_2}a.

The next sequence of optimizations addressed dominantly loop operations. In the \textit{shear} kernel, we removed an intermediate temporary table, trading memory intensity for compute intensity. In the \textit{stream} kernel, we reordered the loops to remove the need for reductions. In both cases, the new code was faster on both the AMD and NVIDIA GPU-based systems, but the overall impact on the CRAY-compiled code for the AMD GPUs was significantly larger. This speedup is shown in Figure \ref{fig1_2}b. We did not investigate if the root cause was compiler optimization or GPU architectural differences.

\begin{figure}[h]
  \centering
  \begin{overpic}[width=\linewidth]{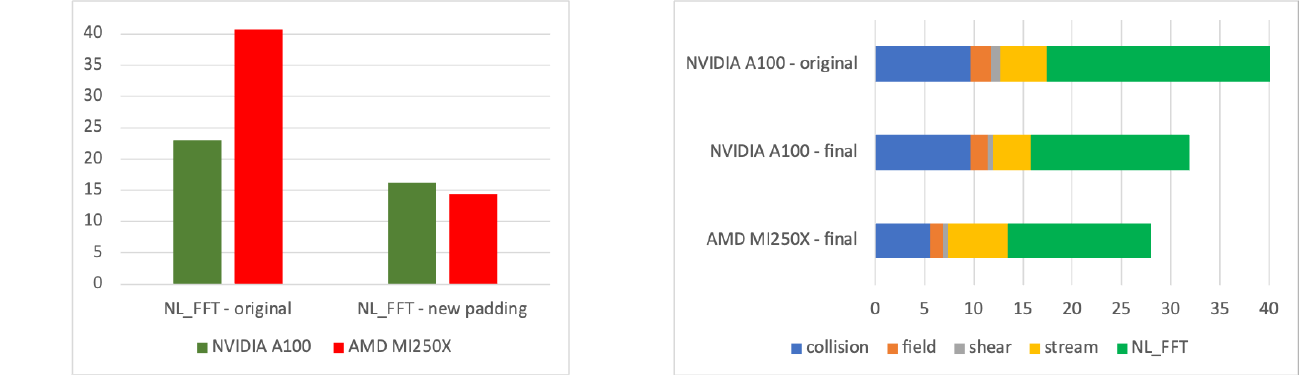}
  \put (6.5,2) {(a)}
  \put (53,2) {(b)}
  \end{overpic}
  \caption{(a) The left plot shows CGYRO \shz\ benchmark results for wallclock time (s) comparing performance of the nonlinear FFT kernel on the NVIDIA A100 (green) and AMD MI250X (red) GPU-based systems for the original padding scheme for dealiasing and the new optimized padding scheme.  (b) The right plot shows the overall performance and timings decomposed into individual CGYRO kernel operations for each system.  "Original" indicates timings before the AMD-porting effort, while "final" indicates after porting and optimization efforts on both systems.  The code is faster, overall, on the AMD GPU system.}
  \Description{Histograms with benchmark results.}
  \label{fig3_4}
\end{figure}

Finally, we looked into the performance difference for the nonlinear FFT kernel. CGYRO logic requires large numbers of 2D complex-to-real FFTs. Since most of the time is spent in the vendor-provided libraries, we could not change the algorithm itself. We thus took a closer look at the input parameters. Like most FFT clients, CGYRO uses zero-padding to avoid aliasing, but this is somewhat flexible. We discovered that the original implementation used a poorly-designed padding scheme that often resulted in very large prime numbers when decomposing in one of the dimensions. By improving the padding scheme to provide decompositions that eliminate large primes, we obtained significant speedups on all platforms. As shown in Figure \ref{fig3_4}a, the improvement was again significantly greater on the AMD GPUs, indicating that the NVIDIA libraries are more tolerant of sub-optimal programming patterns.  Taken all together, CGYRO performance on the AMD GPU-nodes on OLCF Frontier are now faster than on the NVIDIA GPU-based nodes on NERSC Perlmutter for the \shz\ input simulation case, as shown in Figure \ref{fig3_4}b.

\begin{figure}[h]
  \centering
  \begin{overpic}[width=\linewidth]{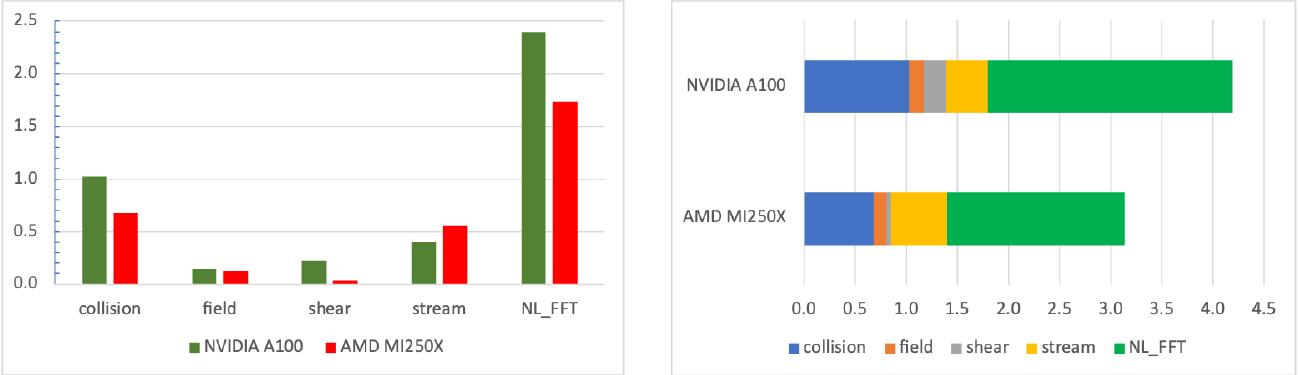}
  \put (1,2){(a)}
  \put (53,2){(b)}
  \end{overpic}
  \caption{(a) The left plot shows CGYRO em04b benchmark results for wallclock time (s) using the "final" optimized code comparing performance of various kernel operations on NVIDIA A100 GPU-based system (green) and the AMD MI250X GPU-based systems (red).  On both systems, CGYRO was run using 288 GPUs. (b) The right plot shows overall performance and timings decomposed into individual CGYRO kernel operations for each system.  The code is overall faster on the AMD GPU system.}
  \Description{Histograms with benchmark results.}
  \label{fig5_6}
\end{figure}

Since CGYRO is most valuable for much larger problems, the performance of the final optimized coded was also compared for the \emz\ test case.  This case represents current leading-edge multiscale simulation sizes.  Since the simulation case operates on a much larger multi-dimensional matrix, it was run on 288 GPUs on both systems. As shown in Figure \ref{fig5_6}, the performance pattern is very similar to \shz, confirming that the optimized code is indeed generally faster on AMD GPUs, as expected. 

Thus, while there are likely still opportunities for improvement, we consider the updated code ready for production use on exascale systems like OLCF Frontier.
Finally, we would like to emphasize that the optimizations benefited performance on both AMD GPU and NVIDIA GPU-based systems, although to a different degree. The unexpected differences in performance were a great lead and motivator for this activity, showing the value of heterogeneity in the HPC arena.

\section{Comparing inter-GPU communication}

Since most CGYRO simulations operate on large memory buffers and require multiple GPUs, communication between GPUs tends to contribute to a large fraction of the total simulation time. Internally, CGYRO splits the multi-dimensional physics problem into a two-dimensional communication problem, with one dimension dominated by all-to-all MPI communication (i.e., global buffer transpose operations), and the other dominated by all-reduce MPI communication \cite{sfiligoi:2022}.  The amount of data exchanged in the first communication dimension is large but is fixed. In the second dimension the amount of data exchanged starts from a much smaller baseline but then grows linearly with size. The ratio between the two dimensions is relatively flexible, with the ideal ratio dependent on the size of the problem.

Both of the systems being compared in this work are composed of nodes with 4 fully-connected GPU chips. The NVIDIA A100 GPUs on NERSC Perlmutter are connected using NVIDIA's NVLINK at 4x 25 GBps, while the AMD MI250X GPUs on OLCF Frontier are connected using AMD's Infinity Fabric at 2x 50 GBps. Both systems use HPE Slingshot 4x 25 GBps networking for inter-node communication. However, the networking endpoint (NIC) setup inside the node is different between the two systems. The NERSC system has 4 NICs on the PCIe bus shared between the CPU and the 4 GPUs, while the OLCF system has one NIC directly attached to each GPU. 

\begin{figure}[h]
  \centering
  \begin{overpic}[width=\linewidth]{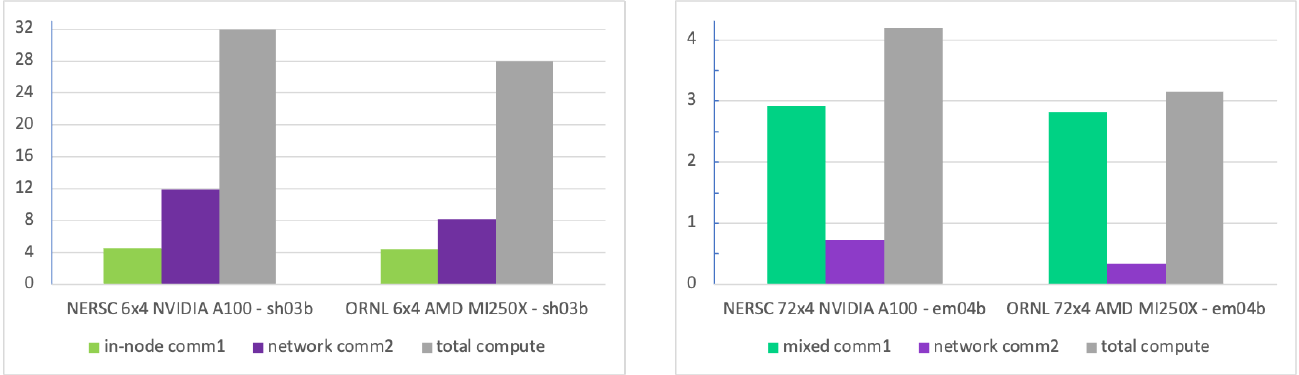}
  \put (1,2){(a)}
  \put (53,2){(b)}
  \end{overpic}
  \caption{CGYRO communication and compute benchmark results for wallclock time (s) using the "final" optimized code comparing performance on the NERSC NVIDIA A100 GPU-based system and the OLCF AMD MI250X GPU-based system.  (a) The left plot shows results for the \shz\ benchmark simulation, using 6 nodes.  (b) The right shows results for the \emz\ benchmark simulation, using 72 nodes.  The dedicated-NIC setup of the OLCF system delivers greater performance than the shared-NIC setup of the NERSC system. }
  \Description{Histograms with benchmark results.}
  \label{figc1_2}
\end{figure}

For the \shz\ benchmark, which used 6 nodes, the first communication dimension was kept fully inside each node, as there is significantly more communication bandwidth available there. With the two dimensions approximately balanced, the data exchanged over the network in the second dimension was still smaller than the data exchanged inside the node. As shown in Figure \ref{figc1_2}a, the two systems showed about the same communication performance inside the node.  This aligns with the hardware specifications. On the network-traversing second dimension, however, the dedicated-NIC setup of the AMD GPU-based system on OLCF Frontier performs significantly better than the shared-NIC setup of the NVIDIA GPU-based system on NERSC Perlmutter.

For the \emz\ benchmark, which used 72 nodes, the same strategy could not be adopted since the ratio would be disproportionately skewed toward the second dimension. Thus,  instead, the first communication dimension was spread as widely as possible; that is, over 36 nodes. All communication then at least partially traversed the network. As shown in Figure \ref{figc1_2}b, the two systems again performed similarly for the first dimension but not for the second.  This indicates that the difference between the NIC setups has less influence on all-to-all MPI traffic than on all-reduce MPI traffic.

Overall, the in-node communication capabilities of AMD GPUs are almost identical to the NVIDIA GPUs. However, the dedicated-NIC setup in the AMD GPU-based OLCF Frontier system delivers greater performance than the shared-NIC setup in the NVIDIA GPU-based NERSC Perlmutter system. That said, the differences account for only a small fraction of the typical overall CGYRO simulation time.

\section{Summary and conclusions}

The CGYRO code, a widely-used fusion simulation tool written in FORTRAN with OpenACC-based GPU acceleration, has been successfully ported from NVIDIA GPU-based systems to AMD GPU-based systems. The porting from the NVIDIA compiler was relatively straightforward using the CRAY compiler on the AMD system, but the performance optimization required more effort and analysis.  These optimizations were related to loop parallelization, memory management, and FFT padding schemes. While the optimizations targeted and primarily improved performance on newer AMD GPU system, some speedup was also achieved on NVIDIA GPU systems. Overall, while there are likely still opportunities for improvement, we consider the updated code ready for production use on the OLCF Frontier exascale system.

\begin{acks}
This work was partially supported by the U.S.\ Department of Energy under awards DE-FG02-95ER54309, DE-FC02-06ER54873, and DE-SC0017992, and by U.S. National Science Foundation (NSF) Grant OAC-1826967. An award of computer time was provided by the INCITE and ALCC programs. This research used resources of the Oak Ridge Leadership Computing Facility, which is an Office of Science User Facility supported under Contract DE-AC05-00OR22725. Computing resources were also provided by the National Energy Research Scientific Computing Center, which is an Office of Science User Facility supported under Contract DE-AC02-05CH11231.
\end{acks}

\bibliographystyle{ACM-Reference-Format}
\bibliography{sample-base}


\begin{thebibliography}{3}


\ifx \showCODEN    \undefined \def \showCODEN     #1{\unskip}     \fi
\ifx \showDOI      \undefined \def \showDOI       #1{#1}\fi
\ifx \showISBNx    \undefined \def \showISBNx     #1{\unskip}     \fi
\ifx \showISBNxiii \undefined \def \showISBNxiii  #1{\unskip}     \fi
\ifx \showISSN     \undefined \def \showISSN      #1{\unskip}     \fi
\ifx \showLCCN     \undefined \def \showLCCN      #1{\unskip}     \fi
\ifx \shownote     \undefined \def \shownote      #1{#1}          \fi
\ifx \showarticletitle \undefined \def \showarticletitle #1{#1}   \fi
\ifx \showURL      \undefined \def \showURL       {\relax}        \fi
\providecommand\bibfield[2]{#2}
\providecommand\bibinfo[2]{#2}
\providecommand\natexlab[1]{#1}
\providecommand\showeprint[2][]{arXiv:#2}

\bibitem[Belli et~al\mbox{.}(2022)]%
        {sfiligoi:2022}
\bibfield{author}{\bibinfo{person}{E.A. Belli}, \bibinfo{person}{J. Candy},
  \bibinfo{person}{I. Sfiligoi}, {and} \bibinfo{person}{Frank
  F.~W\"{u}rthwein}.} \bibinfo{year}{2022}\natexlab{}.
\newblock \showarticletitle{Comparing Single-Node and Multi-Node Performance of
  an Important Fusion HPC Code Benchmark}. In
  \bibinfo{booktitle}{\emph{Practice and Experience in Advanced Research
  Computing}} (Boston, MA, USA) \emph{(\bibinfo{series}{PEARC '22})}.
  \bibinfo{publisher}{Association for Computing Machinery},
  \bibinfo{address}{New York, NY, USA}, Article \bibinfo{articleno}{10},
  \bibinfo{numpages}{4}~pages.
\newblock
\showISBNx{9781450391610}
\urldef\tempurl%
\url{https://doi.org/10.1145/3491418.3535130}
\showDOI{\tempurl}


\bibitem[Candy et~al\mbox{.}(2016)]%
        {candy:2016}
\bibfield{author}{\bibinfo{person}{J. Candy}, \bibinfo{person}{E.A. Belli},
  {and} \bibinfo{person}{R.V. Bravenec}.} \bibinfo{year}{2016}\natexlab{}.
\newblock \showarticletitle{A high-accuracy Eulerian gyrokinetic solver for
  collisional plasmas}.
\newblock \bibinfo{journal}{\emph{J. Comput. Phys.}}  \bibinfo{volume}{324}
  (\bibinfo{year}{2016}), \bibinfo{pages}{73}.
\newblock
\urldef\tempurl%
\url{https://doi.org/10.1016/j.jcp.2016.07.039}
\showDOI{\tempurl}


\bibitem[Candy et~al\mbox{.}(2019)]%
        {candy:2019}
\bibfield{author}{\bibinfo{person}{J. Candy}, \bibinfo{person}{I. Sfiligoi},
  \bibinfo{person}{E. Belli}, \bibinfo{person}{K. Hallatschek},
  \bibinfo{person}{C. Holland}, \bibinfo{person}{N. Howard}, {and}
  \bibinfo{person}{E.D`Azevedo}.} \bibinfo{year}{2019}\natexlab{}.
\newblock \showarticletitle{Multiscale-optimized plasma turbulence simulation
  on petascale architechtures}.
\newblock \bibinfo{journal}{\emph{Computers \& Fluids}}  \bibinfo{volume}{188}
  (\bibinfo{year}{2019}), \bibinfo{pages}{125}.
\newblock
\urldef\tempurl%
\url{https://doi.org/10.1016/j.compfluid.2019.04.016}
\showDOI{\tempurl}


\end{thebibliography}

\end{document}